\documentclass[twocolumn,secnumarabic,amssymb,nofootinbib,superscriptaddress]{revtex4-1}
\usepackage{hyperref}
\usepackage{amssymb}
\usepackage{amsmath}
\def\be{\begin{equation}}
\def\ee{\end{equation}}
\begin{document}
\title{Space-time emergence from individual interactions}
\author{Anna Karlsson}
\email{karlsson@ias.edu}
\affiliation{Institute for Advanced Study, School of Natural Sciences, \\
1 Einstein Drive, Princeton, NJ 08540, USA\\}
\affiliation{Division for Theoretical Physics, Department of Physics, \\
Chalmers University of Technology, SE-412 96 Gothenburg, Sweden}
\begin{abstract}
A scenario of space-time emergence from individual interactions at the quantum level requires correlations encoding a scalar product, for large-scale emergence of space orientation. In \mbox{$D=4$} the candidate is spin 1/2 correlations, indicating complementarity to be crucial in models of correlations, entangling processes and thermalization giving rise to space-time. We argue for the relevance of identifying fundamental interactions and of analysing how they might give rise to space-time emergence. We discuss how to model complementarity locally, contrasting with how it is absent in the wave function due to a restriction to classically simultaneous variables. In modelling theory after measurements, we suggest an extended model to contain `orthogonal' information existing in parallel, in a vector analogy. We also give a rough conjecture of what would be required of entangling processes for equilibration of correlations between vacuum fluctuations, only briefly mentioning effects of acceleration.
\end{abstract}
\maketitle

\section{Introduction}
For space-time to be an emergent feature at the quantum level, it must equivalently originate in individual interactions, since interactions constitute what can be observed and hence what can be reliably modelled on. A clear picture of what the relevant interactions are would be useful for understanding quantum gravity, and so it is of interest to identify and analyse them and how they give rise to space-time. In recent years, entanglement entropy in regions of conformal field theory (CFT), giving rise to gravity through gauge/gravity duality \cite{Maldacena:1997re}, has been widely discussed and entanglement has been recognized as important for space-time emergence \cite{VanRaamsdonk:2010pw}. Focussing on emergent space instead of gravity, one requirement is apparent: for orientation to emerge, the relevant interactions must be able to communicate a scalar product. In $D=4$, the only readily available candidate for this is spin 1/2 correlations. The objective of this text is to argue for the importance of identifying and analysing the different correlations (fundamental interactions) relevant for space-time emergence and how entangling processes may give rise to equilibration between vacuum fluctuations, in a conjecture of thermalization giving rise to the appearance of space-time at large scales. Our key example is spin 1/2 correlations and emergence of space orientation, but conceptually a large part of the discussion carries over to interactions relevant for space-time emergence in general.

Connections between space-time and thermodynamic properties have long been observed \cite{Bekenstein:1973ur,Jacobson:1995ab,Wald:1999xu}. Recently even a model of thermalization of entanglement \cite{Hubeny:2018tah} was proposed, highlighting that interacting systems typically thermalize and that such equilibration is likely to carry over to quantum systems \cite{Short:2012qvq}, providing an emergence of macroscopically perceived entities such as dynamic space-time. While these types of approaches are interesting, including entanglement entropy investigations, the presence of interactions involving complementarity makes it plausible that the individual interactions are of central importance for issues with quantum gravity (e.g. strong interactions). If so, the very assumption of space-time must be dispensed of as a starting point, including space and fields in space-time, and allowed to emerge through entangling processes. The conjecture of thermalization of those interactions then conceptually overlaps with \cite{Hubeny:2018tah} and arguments therein. While the general motivation partly overlaps with \cite{Penrose:1972jq}, our approach through complementarity is different from twistor theory.

Note that metrics $g_{\mu\nu}$ will not be discussed, since the present analysis concerns interactions at a sublevel. We consider what would be required for large-scale emergence of the tension structure a metric describes, analogously to how temperature arises from collisions. Since we defer a discussion of quantization of time, no prescription for a metric can be given. The issue of space is a sufficiently interesting first subject. 

We focus on a requirement for space to emerge locally: the existence of a local scalar product, arising through interactions (\S\ref{s.emerge}). Individual interactions encoding scalar products are currently considered non-local due to the proof of Bell's theorem \cite{Bell:1964kc,Clauser:1969ny}. Our key finding is that the information structure of complementarity (\S\ref{s.comp}) supports a formalism alteration which sidesteps Bell's theorem and provides local EPR correlations (\S\ref{s.beyond}). Complementarity can equally be regarded as information stored orthogonally (like different vector entries). Once non-scalar information is considered, classical probability theory must be extended to a vector concept too, for validity, and this precisely coincides with removing the structure which in Bell's theorem enforces non-locality. Hence our alternative interpretation contains the necessary and sufficient extension for a space-time formulation to be \emph{local}, which is an important concept, making the construction worth considering. We note that the non-locality of Bell's theorem arises from a classical interpretation of quantum entities, both for variables/operators \emph{and} probability theory. How complementary information would be encoded is discussed in detail in \S\ref{s.emerge}. While we use the scalar product as a key example, the vector concept extends to complementary information in general.

As apparent for spin, the behaviour of interactions related to space-time is irrevocably connected to complementarity, for information on position in and movement through space-time is complementary. Indeed, complementarity is as an important feature of quantum physics as the existence of quanta: both are fundamental quantum features distinct from classical physics. However, as identified in the Einstein--Podolsky--Rosen (EPR) paradox \cite{Einstein:1935rr}, the wave function formalism fails to encode complementarity, a limitation which is important to recognize. While entanglement entropy constructions skirt this dilemma, an approach based on individual interactions cannot. To find and motivate a suitable way of modelling the interactions, we begin with a discussion on complementarity in relation to the wave function, before moving on to how to best model complementary relations. Since a restriction to a classical (scalar) setting with simultaneous variables is not compatible with measurements, our conjecture involves information existing in parallel in a vector structure. We discuss how this strictly is required by complementarity, measurements and the EPR correlations, how the principle of locality would be respected, and how spin 1/2 and photon polarization can be well understood in this setting.

We end with a rough conjecture of what thermalization would require. The entangling processes between spin 1/2 correlated particles would have to differ from measurements in not being destructive, sot that local equilibrium configurations can be reached, followed by some thermalization process inferring space on large scales. The finer points would be dependent on effects of time and gravity, which we only briefly comment upon. In total, the complementary sets relevant for space-time emergence ought to be position/momentum, the angular counterparts, spin and time/energy,
\be\label{eq.sp.comp.vars}
(x_i,p_i)\,,\quad(r,L)\,,\quad\{J_i\}\sim(r,\{L_j\})\,,\quad(t,E)\,,
\ee
in $d$ dimensions with $i,j\in\mathbb{Z}^+,\,i\leq d,\,j\leq d-1$. In a scenario of emergent space, two features are crucial: generation of relative \emph{distance} and \emph{orientation}, through fundamental interactions. Position/momenta $(x_i,p_i)$ readily provides the former, with an angular version in $(r,L)$. For the latter, an identification of a scalar product is required, as mentioned above, and spin 1/2 has the only $D=4$ correlations which in pair production encodes multiple directional correlations simultaneously. The alternative representation of a set of $J_i$ in terms of angular position and rotations will be motivated below. For now, we note that spin has a dual in one dimension lower in photon polarization, also described by these representations. Also, while entanglement strictly does not require complementarity, in this text entanglement and entangling processes always refer to correlations and interactions involving complementarity.

\section{Complementarity \& the wave function}\label{s.comp}
Complementarity is a quantum feature which initially was thought of as a disturbance by measurement \cite{Heisenberg1927,Bohr:1928vqa}, but since has been identified as an inherent property. Its nature is conceptually clear in relations like that for position and momentum\footnote{\cite{Heisenberg1927} discusses how this goes beyond the observation that a measurement of one entity infers uncertainty of the other(s).}
\be\label{pdx}
p_i\propto dx_i\,.
\ee
A value of one entity is irrevocably connected to all its complementary variables being undefined, similar to how a change $d x_i$ is required for a value of $p_i$ and incompatible with a value of $x_i$. Complementary variables, e.g. $(x_i,p_i)$, are not simultaneous scalar entities\footnote{Note that vectors in space such as $\vec{p}$ have parts that are simultaneous scalars in this respect. A function of the set of $p_i$: $f(\{p_i\})$ is consistent, while the same is not true for spin. We also discuss complementarity rather than conjugate variables, since sets of complementary variables are not restricted to pairs.}. They cannot be probed to arbitrary accuracy from one and the same particle, nor does it make sense to use them side by side, as scalar entities, in any theoretical model: 
\be\label{fneq}
f\neq f(x_i,p_i)\,. 
\ee
In modelling theory after experiment, this deviation from classical reality has presented a challenge to modern physics. Pair production permits measurements on complementary variables (one per particle), and the pair correlations cause the EPR paradox \cite{Einstein:1935rr}.

The issue concerns what represents physical reality. Classically, physically real entities can be measured simultaneously. In quantum physics, complementarity prohibits this. At a measurement (destructive) at most one complementary variable can be determined. Yet quantum theory states that complementary variables are correlated in pair production, e.g. through conservation of momenta. Correlations of this type are measurable one at a time, and can thus only be statistically inferred on each specific type of pair production. While such statistical correlations do not represent physical reality in the classical sense (and it is a logical fallacy to infer the correlations for single particles) when inferred by measurements, the ensemble behaviour must represent physical reality, by definition. Measurements are what define physical reality.

To clarify the role of complementarity vs. the wave function with respect to physical reality, we now summarise some relevant history. EPR \cite{Einstein:1935rr} highlighted a paradox between two theoretical frameworks: that the quantum theory correlations for complementary variables are not captured by the wave function unless non-local. There are three ways out of that: \emph{(1)} non-locality, \emph{(2)} replacing the wave function, or \emph{(3)} a breakdown of the quantum theory predictions. Either way, causality holds since EPR correlations do not represent signals. To exclude the correlations being a product of system settings at the time of pair production \cite{Bohr:1935af}, decisive experiments also need to have the settings for measurement changed during the flight of the particles. A precise way to analyse the issue \cite{PhysRev.108.1070} is in terms of spin or linear photon polarization correlations,
\begin{subequations}\label{pai}
\be\label{pai.1}
P(a,i)=\frac{1}{2}\,,\quad \left\{\begin{array}{llll}
a\in S^1\,,&i\in\{1,0\}\,,&\text{photon},\\
a\in S^2\,,&i\in\{+,-\}\,,&\text{spin 1/2},
\end{array}\right.
\ee
\be\label{pai.2}
\sum_iP(a,i;b,i)=\left\{\begin{array}{ll}
{\displaystyle(a\cdot b)^2}\,,&\text{photon},\\[1ex]
{\displaystyle\frac{1-a\cdot b}{2}}\,,&\text{spin 1/2},
\end{array}\right.
\ee
\end{subequations}
with measurements in the directions $(a,b)$ on the separated parts of the pair, at $A$ and $B$. Recently, experiments \cite{Hensen:2015ccp,PhysRevLett.115.250401,PhysRevLett.115.250402} showed the actual `in-flight', space-like separated correlations to be within experimental error of \eqref{pai} \emph{and} outside what can be locally captured by the wave function, even if improved by hidden variables, as identified for spin in Bell's theorem \cite{Bell:1964kc,Clauser:1969ny}.

The essential message from right above is that the quantum correlations stand up to experimental tests, on macroscopic scales. The physical reality of their complementarity and statistical correlations is inferred by measurements. Moreover, the wave function, by now nearly synonymous with quantum physics, characterises EPR correlations as causal and non-local. Interpretations of this range from that the wave function fails to capture conditional probabilities concerning complementarity, to the ER=EPR conjecture \cite{Maldacena:2013xja}, equating EPR correlations with Einstein--Rosen bridges \cite{Einstein:1935tc} (wormholes).

Returning to the discussion on theory construction, it is clear from \eqref{pdx} that complementarity is a fundamental feature of \emph{quantum} physics with central conceptual implications. It may well be as important as the existence of quanta, and ought to be a natural part of a complete description of quantum physics. However, it is not present in the wave function formalism. The wave function is a probability distribution for what an observer will encounter, constructed from modern probability theory and measure theory. Conditional probability
\be\label{basis}
\int \mathrm{d}\lambda \,\rho(\lambda) \,\ldots
\ee
(for some probability density $\rho$) is defined for any set of variables $\{\lambda\}$, but relies on this set to be `measurable'. The mathematical term involves countable additivity of disjoint sets. Effectively, the considered variables are required to be simultaneously measurable. Properties which cannot be divided into independent entities, as with complementarity, are disregarded by construction. For example, no set of independent variables can capture the physics of spin in $d=3$ (multiple separate, dependent ones are required) and so the construction \eqref{basis} in Bell's theorem \cite{Bell:1964kc,Clauser:1969ny} is tantamount to leaving out complementary relations. This restriction also is why the wave function only is defined for domains of classically simultaneous variables, with a dual in position/momentum space.

A definition of what is measurable that leaves out complementarity is absurd in the quantum regime. Through pair production, two complementary variables can be measured, and statistical correlations inferred. Here is where the classical definition of measure theory ceases to be valid, and where it needs to be extended to capture the physical reality verified through experiments. \emph{The wave function does not describe quantum physics in the same sense that probability theory restricting to simultaneously measurable variables does not describe complementarity}. Having identified this limitation of the wave function, we conclude that to accurately capture complementarity it is necessary to consider probability theory beyond classical concepts, while compatible with \eqref{fneq} (e.g. different from phase space considerations).

A more pragmatic angle to identifying the shortcomings of the wave function is the following: if a theory displays causal non-local correlations, the non-locality clearly is an artefact of the formulation, since non-locality only can be proven physical through the verification of non-local signals. Since the wave function formalism cannot be altered to locality \cite{Bell:1964kc,Clauser:1969ny} and the measured correlations would require non-locality, a new definition is required. 

The wave function is very useful when complementarity is disregarded. We emphasise that its limitation with respect to complementarity only is clarified through experimental verification of statistical correlations not representing signals, and those correlations appearing non-local due to a restriction to a classical notion of what is measurable, built into the formalism. Since what is measurable differs from this classical notion, it is necessary and possible to formulate a theory with local EPR correlations.

\section{Beyond the wave function}\label{s.beyond}
To model quantum physics with complementarity, a construction different from the wave function formalism is required. The physical reality of complementarity includes \emph{(1)} not simultaneously measurable variables and \emph{(2)} pair production with statistical correlations (causal and space-like separated) between complementary variables, which cannot be locally described in the absence of \emph{(1)}. The argument is circular when only simultaneously accessible variables are considered, so the logical solution is to extend the concept of information from a scalar entity to multiple entities existing in parallel. Here, the clearest analogy is that of vectors (not in space-time). As $(x_i,p_i)$ are not simultaneous scalar entities, nor is the information pertaining to them, which rather is mutually orthogonal, existing in parallel, and only accessible to an observer through a single projection (per particle). In this analogy, one complementary variable can be fully measured or multiple ones be partially accessed, in compliance with the uncertainty principle \cite{Heisenberg1927}. Meanwhile, entangling processes represent simultaneous interactions in multiple channels, and pair production gives one-to-one correlations in each channel of information, i.e. full entanglement. Complementarity is reinterpreted as orthogonality of information, with complementary variables simultaneous and orthogonal to each other in terms of accessibility of the information they represent. While unorthodox, to our knowledge nothing prohibits this type of consideration.

A second conceptual change is that a model with information encoded in parallel represents a system formulation, whereas the wave function describes the observer experience, where a classical probability setting is intuitive. Since theory strictly only is determined by what is measured, it is equally viable to aim for an objective model of the system instead. While necessary for complementarity, this change has further consequences. Modelling on the observer experience effectively puts the observer at the centre of the universe, a choice entailing more artefacts than the EPR paradox, e.g. that of Schr\"odinger's cat. As much as the wave function is a prediction of what the observer will encounter, it is also a statement on lack of information on behalf of the observer. Without previous interaction, a foreign subsystem appears undetermined, without necessarily being so except to the observer in question. However, while many pardoxes are of a philosophical nature, locality is not. Hence, it is interesting that complementarity requires an objective formulation.

An objective picture with information stored in parallel opens up for several different formulations, beyond the scope of this text. Focussing on the spin/photon pair correlations in \eqref{pai}, we conjecture the complementarity formulation to simply be \eqref{pai} with
\be\label{pai.ext}
a\cdot b \rightarrow a\big|_A\cdot b\big|_B\,.
\ee
Here, locality is made possible through the consideration of ($d$) multiple channels of orthogonal information. \eqref{pai} represents the simplest objective observation to make, without any specification of the individual systems prior to measurement, yet capturing the relations inferred by pair production. Meanwhile, \eqref{pai.ext} includes an assumption of acceleration to change the notion of orientation at particle level, so that the pair correlations accurately capture relative curvature between $(A,B)$.

Reconnecting to emergence of space-time, the complementarity picture furnishes a way to investigate how space-time might arise from individual interactions. The scalar product in the spin/photon correlations single them out as candidate origins of emergence of \emph{orientation}. However, a first question must be how to understand spin and photon polarization. The general argument of parallel information accommodates for the correlations, but gives little explanation of their characteristics. Below, we will discuss how to best understand both the individual spin/photon pair correlations and the limits to what can be measured, as well as what might be involved in an emergent picture.

\section{Relative orientation from pair correlations}\label{s.emerge}
Emergent directions require an identification of a scalar product at the level of individual interactions,
\be\label{svp}
\sum_{i=1}^d |\hat e_i\rangle\langle\hat e_i| \,\,\,\, \longleftrightarrow \,\,\,\,\frac{1}{N}\sum_{n=1}^N f(a_n,b_n) \xrightarrow{N\rightarrow\infty} a\cdot b\,,
\ee
for $\forall\, a_n\cdot b_n=a\cdot b$, or something corresponding to the rhs in a less idealized setting. At the quantum level, this abstract connection $f(a_n)$ may also be expected to give output in terms of quanta, i.e. discrete values at each site of measurement $(A,B)$. On the lhs, the scalar product shows the classically counterintuitive nature of this type of connection: parallel, simultaneous correlations through multiple channels, where classically only one at a time is possible. Correlations of this type are required to be complementary, and the only candidates (identified so far) are spin 1/2 and linear photon polarization entanglement, since their pair correlations \eqref{pai} represent versions of \eqref{svp}. For spin, the uncertainty relation $[J_a,J_b]\propto (a\times b)$ also illustrates the overlap in information of $a\cdot b$. 

The central role of the spin 1/2 and photon polarization correlations makes it desirable to understand their general characteristics and similarities. Importantly, they are directional in nature (probed at an angle), with non-trivial rotation symmetry. It turns out to be useful to describe them through a (rotation symmetric) representation $(r,\{L_i\})$, with an angular position $r$ and a set of angular rotation vectors $L_i$, in total giving $d$ operators. While this basis represents intrinsic qualities, not literal rotation, the structure is analogous in terms of conservation and elucidates the correlations, the limitations to what is measurable, the connection to a scalar product and the duality of the $2$ and $3d$ settings, i.e. spin vs. photons.

Beginning with the simpler case of $d=2$ (photons), the correlations\footnote{The circular photon polarization is set apart from the EPR correlations. With $j\in\{+,-\}$ along the direction of propagation, $P(j,A)=P(j,A;j,B)=1/2$. The circular--linear correlation is trivial: $P(j,A;b,i)=1/4$.} are reproduced by rotation matrices in the plane perpendicular to the direction of propagation ($\vec{p}$), with a symmetry of $\varphi\sim\varphi+\pi$. With a reference direction $r$ and positive orientation set by a direction of rotation $L=\pm\vec{ p}/|p|$, a measurement can be given relative to the internally defined reference frame as
\be\label{photon.a}
\pm a \,\,\rightarrow\,\, \mathfrak{a}=R(\varphi_a)\hat r=\begin{bmatrix}\cos\varphi_a&-\sin\varphi_a\\\sin\varphi_a&\cos\varphi_a\end{bmatrix}\begin{bmatrix}1\\0\end{bmatrix}\,,
\ee
with $\varphi_a\in [0,\pi)$. The correlations are then captured by
\be\label{papb}
m_\mathfrak{a}=R(\varphi_a)\begin{bmatrix}1&0\\0&0\end{bmatrix}R(-\varphi_a)\,:
\quad \text{tr}(m_\mathfrak{a}m_\mathfrak{b})=(a\cdot b)^2\,,
\ee
in an overlap picture that contains no way of assigning a definite outcome at either site $(A,B)$, giving \eqref{pai.1} as a consequence of the randomness of $r$.

That a pairwise shared\footnote{For photons, a shared $L$ means $L\big|_A=\hat{\vec{p}}\big|_A\Rightarrow L\big|_B=\hat{\vec{p}}\big|_B$.} $(r,L)$ captures the correlations is illuminating in terms of what can be determined, and for the duality of circular/linear polarization. $L$ literally represents circular polarization evolving upon interaction, while the linear outcome depends on a combination of $(r,L)$. Of these, only one can be fully determined at a time. Selecting on $L$ for one half of the pair would, at the other end, give a fully correlated outcome for circular polarization, and random results for the linear case. In selecting on linear polarization instead, $L$ remains undetermined but present in the rotational correlations, equivalently posed in terms of one shared, undetermined $L$ and an angular distance between measurement angles. The setting is equivalent to $e^{2i\varphi}$, which in terms of $e^{2i\theta}e^{2i\varphi_a}$ aptly illustrates the further correlations required due to that a vector $a$ is not uniquely defined by its components squared\footnote{Models on these typically require negative probabilities, to correct overestimated correlations.} $\{a_i^2\}$. These correlations describe a classically unorthodox mixing of relative probabilities.

The assignment of a definite value is in turn equivalent to the unitary outcome of $e^{2i\varphi}$ (norm $1$) with a simultaneous parallel assignment of values\footnote{To see this requirement, consider what is required for consistent assignments for different $\varphi$, connected by rotations. For the explicit decomposition in \eqref{e.multiple}, recall that $\cos2\varphi=\cos^2\varphi-\sin^2\varphi$, and the corresponding for $\sin2\varphi$.} in the real and imaginary channels
\begin{gather}\begin{aligned}\label{e.multiple}
&e^{2i\varphi}=\cos 2\varphi+i\sin2\varphi\\
&\Leftrightarrow\quad\left\{\begin{array}{ll}1:\,\cos^2\varphi\,,&-1:\,\sin^2\varphi\\ 1:\,\cos^2(\varphi-\pi/4)\,,&-1:\,\sin^2(\varphi-\pi/4)\end{array}\right.
\end{aligned}\end{gather}
exactly the requirement for an emergent scalar product. Note that the $\pi$ periodicity translates into orthogonal axes in the $2d$ plane at $\varphi\in\{0,\pi/4\}$. Again, the assignment of a definite value is complementary (`orthogonal') to the relative correlation, and so both cannot be simultaneously discussed --- they are only available through pair production.

In the light of the above, the $2d$ picture is clear, and in higher dimensions the construction can be extended to $(r,\{L_j\})$ with a set of angular rotations spanning the unit sphere $S^{(d-1)}$ in a set order. In addition, while pair produced photons share the same $(r,L)$, spin 1/2 particles require opposite characteristics. However, in $d>2$, models of the correlations like \eqref{papb} are not tractable. In $d=3$ two disparate, non-commutative rotations are present and remain undetermined, including the \emph{order} of rotation. The outcome depends on two complementary variables, neither of which can be eliminated or further specified. Here, the rotational origin only gives the observed trigonometric dependence of \eqref{pai.2}. In addition to a scalar product, spin 1/2 and $3d$ orientation also has an observed $4\pi$ rotation symmetry. While how this arises is not apparent (and desirable to understand) the set of two $L_i$ is suggestive of that $2\pi$ rotations are required for each, both in terms of acceleration by rotation for individual particles, and the entangling process discussed below.

As such, a $d=3$ rephrasing of $\{J_i\}$ into $(r,\{L_i\})$ mostly illuminates the complementary qualities of spin (in relation to measurement, same as for photons) and the connection to a scalar product, in $3d$ with correlations through $(\hat x,\hat y, \hat z)$ in a cartesian coordinate system, instead of \eqref{e.multiple}. In addition, it shows a duality of spin 1/2 and photon polarization, similar in their nature in terms of $(r,\{L_i\})$. Allowing for the different dimensionality and the opposite statistics (anti-/correlation), the different periodicity of the internal systems $(\pi,2\pi)$ translates into that the $3d$ correlations are dual to the $2d$ relation with a doubled period ($\pi\rightarrow 2\pi$),
\be
\mathfrak{a}\xrightarrow{\varphi_a\rightarrow2\varphi_a}\tilde{\mathfrak{a}}\,: \quad (\mathfrak{a}\cdot\mathfrak{b})^2= \frac{1+\tilde{\mathfrak{a}}\cdot\tilde{\mathfrak{b}}}{2}\,.
\ee
The dual nature of the correlations makes it plausible that space-time, with $3d$ space orientation emerging from spin 1/2 entanglement, in certain geometries would be dual to $2d$ gauge theory, as in gauge/gravity duality.

\section{Emergence of space orientation}
With pair correlations encoding relative orientation identified, the next requirement for emergence of (large-scale) space orientation is entangling processes besides pair production. For a local equilibrium to be reached, the entities must readily entangle with each other, and the result from two particle interactions must retain information of the initial configurations. The former certainly is true for spin, which also is what is required for $3d$ geometry ($2d$ geometry naturally is restricted to notions of rotation and parity). 

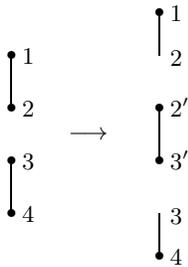
\begin{figure}[tbp]
{\begin{center}
\setlength{\unitlength}{0.8pt}
\begin{picture}(70,115)(0,-95)
\put(0,-5){\circle*{4}}
\put(-0.1,-5){\line(0,-1){25}}
\put(5,-9){$1$}
\put(5,-34){$2$}
\put(0,-30){\circle*{4}}
\put(0,-55){\circle*{4}}
\put(-0.1,-55){\line(0,-1){25}}
\put(5,-59){$3$}
\put(5,-84){$4$}
\put(0,-80){\circle*{4}}
\put(27,-45){$\longrightarrow$}
\put(70,15){\circle*{4}}
\put(69.9,15){\line(0,-1){20}}
\put(75,11){$1$}
\put(75,-10){$2$}
\put(75,-34){$2'$}
\put(70,-30){\circle*{4}}
\put(70,-55){\circle*{4}}
\put(75,-59){$3'$}
\put(69.9,-100){\line(0,1){20}}
\put(75,-85){$3$}
\put(75,-104){$4$}
\put(70,-100){\circle*{4}}
\put(69.9,-30){\line(0,-1){25}}
\end{picture}
\end{center}}
\caption{Illustration of an entangling process between two particles, previously entangled with two other particles. An emergence of orientation through spin entanglement would require the new system to have $(r,\{L_i\})'$ symmetrically dependent on the two initial configurations, with randomness only through how those two initial systems differ, providing interactions towards an equilibrium.\label{f1}}
\end{figure}

Focussing on $3d$ and spin, in comparison a measurement represents a destructive process, altering the complementary qualities. But measurements are also projective interactions, i.e. not entangling, and it is reasonable to believe entangling processes to be of an averaging, stabilizing kind. At least, for space to emerge through interactions, with agreement of orientation on a larger scale, spin is the best candidate so far. 
Figure \ref{f1}
gives a rough illustration of the general idea. Two particles, each entangled to third party entities, entangle into a shared state equally dependent on the two initial figurations, with the outcome fixed if the two initial configurations already are fully entangled, and otherwise distributed in-between the two initial configurations with some probability, giving equilibration instead of copies. Here, `in-between' would be determined by the relative initial configurations. For spin, the easiest approach is in terms of $\{J_i\}$ with a conjectured selection on pairs maximizing $|J_{1,i}\cdot J_{2,j}|$ and a new probability distribution of $J'_{k}$ within each such interval, while accommodating for the orthogonality of $J'_k$. Here, changes $\pm \{J_i\}$ would also have to be considered within the same equivalence class. To the purpose of this text, the precise reassignment is not central --- instead, we focus on the presence of an entangling process furnishing interactions which lead to equilibrium configurations. A suitable overall description might require different parts, such as how local equilibrium is reached (possibly through something like tensor networks) followed by a hydrodynamic formulation.

The general conjecture, as such, entails local pair correlations to entangle in multiple stages and produce an agreement on orientation on a large scale, with smooth changes over large distances. On a local scale, such a structure would have to be supported mostly by vacuum fluctuations, and given boundary conditions by stationary matter (pure geometry). Here, the lifetime of the fluctuations must be greater than the local equilibration time
\be
\tau>\tau_{eq}
\ee 
for the fluctuations to encode any structure. For example, this type of construction could provide a notion of straight lines to moving particles through interactions with their momenta $\vec{p}$, e.g. giving an explanation of the double-slit experiment (particle/wave duality) in terms of geometry through stabilised entanglement exchange through and around the slits. Since any equilibrating process in $3d$ must be crucially affected by gravity, the equilibration is expected to be more complex than discussed above, requiring further analysis beyond the scope of this text. We will merely add some brief comments in relation to time and gravity below. As stated above, the objective of this text is to argue for the relevance of identifying and analysing the different interactions relevant for space-time emergence, through making an example of spin 1/2. A better understanding of all of these interactions is necessary for a more precise understanding of the thermalization, e.g. in terms of something corresponding to a Boltzmann equation.

In modelling emergence of time, the fundamental interaction of relevance most likely is $(t,E)$ entanglement, communicated through interactions. Since acceleration implies time dilation, or a change of some intrinsic definition of a time period $T$ (in addition to length), time and space variations (gravity) might be connected through taking variations in $T$ to define orientation entanglement equilibrium configurations. However, in this framework time would arise both from internal and external causes, anything representing change (i.e. interactions), not quite the type of `clock time' (internal or external) commonly discussed, e.g. in \cite{Page:1983uc}. For the specific example of black holes, there might be a breakdown of space-time structure in close proximity to the horizon. If the individual interactions play a central role near the horizon, instead of the normally present effective picture of a thermalized system, a model of the physics involved would be incompatible with an a priori definition of space-time, while also including strong interactions in the sense of complementarity. The precise effects would depend on the quantum model involving emergence of time. Considering that the relative rate of interaction within a subsystem decreases as it approaches the horizon, time as emergent from individual interactions might give very unorthodox effects, such as a boundary akin to the Zeno paradox of Achilles and the tortoise (black holes `frozen' in time).

\section{Summary}
In emergence of space-time at the level of quantum interactions, complementarity is a crucial part of the quantum physics. An accurate formulation (beyond scalar entities like entanglement entropy) requires inherent complementarity in addition to quantum features, and for this it is necessary to go beyond the wave function and consider information not as restricted to a scalar setting, but existing in parallel. Considerations of this type allows for EPR locality and for intuitive explanations of limitations and results of measurements for e.g. spin and photon polarization, as we have shown. This makes the information structure here advocated, in terms of complementary entities and an extended concept of probability theory, of interest and worth considering: the question of locality is a crucial part of physics.

The construction also allows for conjectures of how the pair correlations, encoding the scalar product required for space, through entangling processes may give rise to a thermalization process resulting in space-time. While our discussion makes an example of emergent space orientation and spin 1/2 correlations, it is relevant for complementary correlations in general, and the approach is of interest for further analysis in terms of effects due to time and gravity. An identification of the relevant fundamental interactions, their entangling properties and how they (possibly) result in a thermalization process might give a better understanding of quantum gravity.

\begin{acknowledgements}
This work is supported by the Swedish Research Council grant 2017-00328. Its initiation was supported by the Knut and Alice Wallenberg Foundation.
\end{acknowledgements}

\providecommand{\href}[2]{#2}\begingroup\raggedright\endgroup
\end{document}